\documentclass[journal]{IEEEtran}
\usepackage[utf8]{inputenc}
\usepackage[T1]{fontenc}
\usepackage[french,main=english]{babel}

\usepackage{cite}
\usepackage{amsmath,amssymb,amsfonts}
\usepackage{algorithm}
\usepackage{algpseudocode}

\usepackage{textcomp}
\usepackage{xcolor}
\usepackage{verbatim}
\usepackage{eurosym}
\usepackage{amsmath}
\usepackage{nccmath}
\usepackage{mathrsfs}
\usepackage{mathtools, bm}
\usepackage{amssymb, bm}
\usepackage{comment}

\usepackage{subcaption}

\usepackage{enumitem}
\usepackage{amsfonts}
\usepackage{relsize}
\usepackage{ragged2e}
\usepackage[free-standing-units=true]{siunitx}
\usepackage{url}

\usepackage{array}
\usepackage{multirow}

\usepackage{tikz}
\usetikzlibrary{positioning}
\usepackage{pgfplots}
\usepackage{graphicx}
\pgfplotsset{compat=newest}
\usetikzlibrary{shapes,arrows,fit,positioning,calc}
\usetikzlibrary{plotmarks}
\usetikzlibrary{decorations.pathreplacing}
\usetikzlibrary{patterns}
\usetikzlibrary{chains,arrows,shapes,spy}
\usetikzlibrary{automata}

\usepackage{textcomp}
\usetikzlibrary{shapes,arrows,backgrounds}
\usetikzlibrary{trees}

\usepackage{siunitx}

\usepackage{soul}
\usepackage{color}
\usepackage{ulem}

\def\BibTeX{{\rm B\kern-.05em{\sc i\kern-.025em b}\kern-.08em
    T\kern-.1667em\lower.7ex\hbox{E}\kern-.125emX}}

\begin{document}

\title{Enhanced Multiuser CSI-based Physical Layer Authentication Based on Information Reconciliation
}
\author{Atsu Kokuvi Angélo Passah, Arsenia Chorti, and
Rodrigo C. de Lamare \vspace{-1.5em}%
\thanks{Atsu Kokuvi Angélo Passah is with ETIS Laboratory, ENSEA, CY Cergy Paris University, CNRS, France, and with the Pontifical Catholic University of Rio de Janeiro (PUC-Rio), Brazil.}%
\thanks{Arsenia Chorti is with ETIS Laboratory, ENSEA, CY Cergy Paris University, CNRS, France, and with Barkhausen Institut gGmbH, Germany.}%
\thanks{Rodrigo C. de Lamare is with the Pontifical Catholic University of Rio de Janeiro (PUC-Rio), Brazil, and with the School of Physics, Engineering and Technology, York University, United Kingdom.}
}

\maketitle

\begin{abstract}
This paper presents a physical layer authentication (PLA) technique using information reconciliation in multiuser communication systems. A cost-effective solution for low-end Internet of Things networks can be provided by PLA. In this work, we develop an information reconciliation scheme using Polar codes along with a quantization strategy that employs an arbitrary number of bits to enhance the performance of PLA. We employ the principle of Slepian-Wolf coding to reconcile channel measurements spread in time. Numerical results show that our approach works very well and outperforms competing approaches, achieving more than $99.80\%$ increase in detection probability for false alarm probabilities close to $0$.
\end{abstract}


\section{Introduction} 
\IEEEPARstart{T}{he} advent of novel wireless systems such as large-scale, heterogeneous, Internet of things(IoT) networks \cite{ref_IoT}, introduces numerous security concerns. In this context, standard cryptographic schemes using public key encryption for node authentication may be broken by quantum computers while post-quantum alternatives may be inappropriate for simple devices due to their computational complexity, or significant delays in low-latency systems \cite{ref3},\cite{ref4}. Therefore, physical-layer security alternatives are of great interest \cite{ref1}, \cite{srmax}, \cite{ref__}. In particular, physical-layer authentication (PLA) protocols are expected to play a crucial role by providing alternative low-complexity and low-latency quantum-resistant solutions. CSI-based PLA is a two-step procedure, including an enrollment (off-line) phase and an authentication (on-line) phase. Typically, higher layer authentication protocols are assumed to be used during the enrollment phase, during which a CSI (or a RF fingerprint) baseline observation is recorded for a node of interest. In the authentication phase, hypothesis testing is subsequently used to verify the consistency of the recorded observation against a new measurement of the CSI.

Recently, several channel-based PLA schemes have been investigated. The study in \cite{ref6} developed an authentication scheme using the channel impulse response (CIR) and incorporated extra multipath delay characteristics of the wireless channel into the authentication framework. Furthermore, this study employed a two-dimensional quantization method to reduce the impact of random variations in amplitudes and delays. Differently from \cite{ref6}, the authors in \cite{ref7} proposed two PLA schemes based on CIR without employing a quantization algorithm. The approach in \cite{ref7} eliminates the quantization errors that can negatively affect the authentication performance. Unlike the other works, \cite{phase} proposed a key-based PLA by exploiting the channel phase response to perform the authentication. In \cite{ref_vtc2024}, preliminary results of a PLA scheme using information reconciliation with a one-bit quantization scheme have been reported for a single-user communication network in the presence of a naive active attacker.

 In this work, we present an information reconciliation scheme based on polar codes along with a quantization strategy that employs an arbitrary number of bits in order to enhance the performance of channel state information (CSI) based PLA in multiuser systems. In particular, we employ the principle of Slepian-Wolf coding reconciliation to reconcile channel measurements spread in time. An analysis of the proposed reconciliation scheme is carried out in terms of probabilities of detection and false alarm. To the best of our knowledge, apart from our preliminary results in \cite{ref_vtc2024} for a single-user system and a scheme with 1-bit quantization, it is the first time reconciliation is being considered in PLA, despite the use of similar concepts in the form of fuzzy extractors in physical unclonable functions and bio-metrics based authentication schemes.

The rest of this paper is organized as follows. Section II presents the system model and explains the authentication phases. In Section III, the proposed approach is described in detail, while performance analyses are carried out in Section IV by deriving closed-form expressions of the probabilities of false alarm and detection. Simulation results are presented in Section V and the paper is concluded in Section VI.

\section{System Model}
\begin{figure}[t]
     \centering    
     \includegraphics[width=0.28\textwidth]{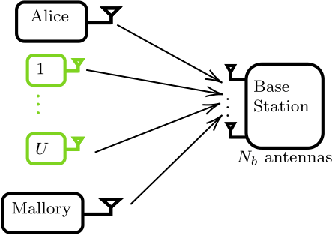}
    \caption{Multiuser system with interference}
     \label{fig1}
\end{figure}
A multiuser wireless communication network is considered Fig. \ref{fig1}. The network includes legitimate nodes, referred to as Alice, Bob (base station) and $U$ legitimate users that act as interfering users during the transmission. In this network, Bob wants to authenticate the user of interest Alice in the presence of the other legitimate users and an adversary Mallory that is a naive active attacker that attempts to impersonate Alice without the use of any precoder or other pre-processing technique \cite{ref_prec}. The objective is to design a scheme based on CSI to distinguish Alice from Mallory in the presence of interfering users. Each user is equipped with a single antenna and Bob is equipped with $N_b$ antennas. 

To achieve a good separation between different users, we assume that the communication occurs in a rich scattering environment and the distance among users exceeds half of a wavelength. Thus, channel attributes between different transmitter-receiver pairs are spatially uncorrelated \cite{jakes}, \cite{goldsmith}. We are testing the proposed algorithm on real datasets and preliminary results look promising but are left as future work.

The authentication comprises two phases, an enrollment phase and an authentication phase. First, in the enrollment phase, Bob estimates Alice's CSI $\mathbf{h}_{a}(t) \in \mathbb{C}^{1 \times N_b}$ in time slot $t$. This offline step allows Bob to use $\mathbf{h}_{a}(t)$ as a point of reference during the authentication phase to determine if the user is indeed Alice or not. Next, during the authentication phase in a subsequent time slot $t+\ell$, Bob takes new CSI measurements of $\mathbf{h}_{u}(t+\ell) \in \mathbb{C}^{1 \times N_b}$, $u\in \{a,m\}$, that may either come from Alice or Mallory, $a$ and $m$ denote, respectively, Alice and Mallory. In this online phase, Bob needs to make an authentication decision based on the CSI obtained in the enrollment phase. We consider a scenario where the CSI changes slowly over time. The channel between the same transmitter-receiver pair can then be well described by 
a first order Gauss-Markov process\cite{ref7}. The channel between Alice and Bob in time slot $t+\ell$ can therefore be expressed as
\begin{equation}
\mathbf{h}_{a}(t+\ell) = \beta\mathbf{h}_{a}(t+\ell-1) + \sqrt{1-\beta^2}\mathbf{n}_a , 
\end{equation}
where $\beta$ is the channel correlation coefficient and $\mathbf{n}_a$ is a measurement noise, ${n_a}_{i} \sim \mathcal{C N}\left(0, \sigma_h^2\right)$, $i = 1, \ldots, N_b$. $\mathbf{n}_a$ is statistically independent of $\mathbf{h}_{a}$.

\section{Proposed Authentication Scheme} 
\begin{figure}[t]
     \centering
    \includegraphics[width=0.45\textwidth]{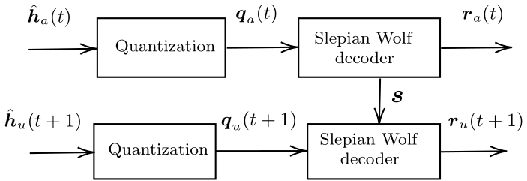}
    \caption{Proposed PLA scheme, $u\in \{a,m\}$} 
     \label{fig2}
\end{figure}

We present a reconciliation scheme to mitigate the impact of disparities in the CSI observed in different time slots. We use the principle of Slepian-Wolf decoding\cite{ref8} to reconcile discrepancies in the CSI measurements over time, as depicted in the system model illustrated in Fig. \ref{fig2}. 
We propose this channel-based PLA scheme that employs quantization with an arbitrary number of bits for multiuser systems. Each phase involves quantizing the CSI, with the output vectors at time $t$ and $t+\ell$ treated as dithered codewords at the input of the reconciliation decoder (Fig. \ref{fig2}). The reconciliation decoder outputs one reconciled vector at each time instance. Note that to this end, the helper data $\mathbf{S}$ (e.g., in the form of a syndrome) generated in the first phase is used by Bob in the authentication phase to reconcile the newly obtained CSI at time $t+\ell$ to the previous one at time $t$. Then, a hypothesis testing is performed to identify the legitimate user versus the impersonator.
\subsection{PLA phases}
The PLA phases are structured as follows.
\subsubsection{Enrollment phase} 
In this offline phase, Bob measures the CSI of Alice as a reference for the authentication:
\begin{equation}
    \centering
\hat{\mathbf{h}}_{a}(t)=\mathbf{h}_{a}(t)+\mathbf{z}(t),
\label{equ2}
\end{equation}
where $\mathbf{z}(t) \in \mathbb{C}^{1 \times N_b}$ is a zero mean complex Gaussian noise so that $z_{i}(t) \sim \mathcal{C N}\left(0, \sigma_z^2\right)$, $i = 1, \ldots, N_b$. $M$ samples of $\hat{\mathbf{h}}_{a}(t)$ are concatenated and by considering the real and imaginary parts, we get the vector $\mathbf{x}_a(t) \in \mathbb{R}^{1 \times N}$ where $N = 2MN_b$. $\mathbf{x}_a(t)$ is then quantized as $\mathbf{q}_a(t)$. 

\subsubsection{Authentication phase}
Without loss of generality, we assume $\ell=1$. In this online phase at time $t+1$, new channel measurements associated with Alice are given by  
\begin{equation}
    \centering
   \hat{\mathbf{h}}_{a}(t+1) = \mathbf{h}_a(t+1) + \sum_{i=1}^{U}\alpha_i^a\mathbf{h}_i^a(t+1) + \mathbf{z}_a(t+1)
    \label{eq3}
\end{equation}
 such that $\mathbf{h}_{a}(t+1) = \beta\mathbf{h}_{a}(t) + \sqrt{1-\beta^2}\mathbf{n}_a$, whereas the new channel measurements of Mallory are described by
\begin{equation}
    \centering
    \hat{\mathbf{h}}_{m}(t+1) = \mathbf{h}_m(t+1) + \sum_{i=1}^{U}\alpha_i^m\mathbf{h}_i^m(t+1) + \mathbf{z}_m(t+1),
    \label{eq4}
\end{equation}
where $\mathbf{z}_a(t+1)$ and $\mathbf{z}_m(t+1)$ are Gaussian noise vectors, $\mathbf{h}_i^a(t+1)$ and $\mathbf{h}_i^m(t+1)$ are interfering terms with interference weights $\alpha_i^a$ and $\alpha_i^m$ respectively. The weights account for the fact that in the channel estimation, different users might use different pilot sequences that can be quasi-orthogonal\cite{pilot}. Similarly to the previous phase, $M$ samples of the channel measurements are concatenated in a vector $\mathbf{x}_u(t) \in \mathbb{R}^{1 \times N}$, $u\in \{a,m\}$, and quantized as $\mathbf{q}_u(t+1)$.

\subsubsection{Quantization}
We use the Lloyd-Max quantizer that is a powerful tool for designing optimal quantizers. 
As $99.7\%$ of the probability density of a Gaussian distribution $(\mu,\sigma^2)$ lies within $\mu-3\sigma$ and $\mu+3\sigma$, the design steps are described by
\begin{itemize}
    \item $n$-bit quantizer $\implies$ $L = 2^n$ quantized levels 
    \item Divide the range $R = (\mu+3\sigma) - (\mu-3\sigma) = 6\sigma$ in $L$ equal intervals 
    \item Calculate each quantizer levels as midpoint of the corresponding interval
    \item Optimize the quantizer levels using the Lloyd-Max algorithm and then store them in the dictionary $\mathcal{D}$
    \item Then the squared error distortion is considered to quantize each sample according to
    \begin{equation}
        \hat{x}_i = \underset{d \in \mathcal{D}}{\arg \min }(x_i-d)^2\text{,}
        \label{eq5}
    \end{equation}
    where $x_i$ is the sample to be quantized and $\hat{x}_i$ is the output.
    \item Codebook: mapping of $\hat{x}_i$ into bits using Gray code. Each Gray code corresponds to a quantized level in $\mathcal{D}$
\end{itemize}

\subsection{Reconciliation}

\subsubsection{Reconciliation} The quantized vectors $\mathbf{q}_a(t)$ and $\mathbf{q_u}(t+1)$ $\in \{0,1\}^{1 \times nN}$, fed into the input of the reconciliation, are mapped to the decoder outputs $\mathbf{r}_a(t)$ and $\mathbf{r}_u(t+1)$, where $nN$ is the codelength. The principle of our reconciliation scheme is based a decoding process using the principle of Slepian-Wolf decoding as in \cite{ref9}, where $\mathbf{q}_a(t+1)$ is decoded using the side
information (helper data) $\mathbf{S}$ derived from the code design and the quantized vector $\mathbf{q}_a(t)$ from the enrollment phase at time $t$. This approach ensures that the correlation between the CSI in both phases, is properly exploited. As a result, it is possible to distinguish Alice from Mallory. Specifically, when attempting to decode Mallory's CSI using the side information, the lack of correlation between their channels prevents Mallory’s CSI from producing similar outputs.
$\mathbf{q}_a(t)$ and $\mathbf{q}_a(t+1)$ are therefore considered as dithered versions of the same codeword. The aim here is to correct measurements errors in order to separate Alice from Mallory, i.e., although reconciliation will be successful for Alice, it will not be successful for Malory. Polar codes \cite{ref9} are the family of error-correction codes examined in this work for reconciliation. Thus, the scheme needs to be designed in such a way that the reconciliation gives (i) in the normal case, i.e,  $\mathbf{q}_a(t)$ and  $\mathbf{q}_a(t+1)$, identical or very close outputs, and (ii) in the spoofing case, i.e,  $\mathbf{q}_a(t)$ and  $\mathbf{q}_m(t+1)$, very different outputs. This also depends on the correlation coefficient and the noise variance.

\subsubsection{Polar codes} 
To enhance the performance of the polar code in finite blocklengths, $\mathbf{q}_u(t+1)$ is decoded using cyclic redundancy check ($\mathrm{CRC}$) successive cancellation list decoding\cite{ref8},\cite{ref10} where the list size is denoted by $L_s$. During the decoding, Bob actually tracks $L_s$ decoding paths simultaneous, where the decoder picks the most likely codeword which satisfies the CRC condition among the $L_s$ paths. Therefore, Bob uses the side information $\mathbf{S}$ that contains the syndrome (frozen bits) and the CRC bits. The $\mathrm{CRC}$ assists the decoder in choosing the right decoding path from a list of possibilities. This syndrome is generated using the quantized vector $\mathbf{q}_a(t)$ and the reliability sequence used for the polar code construction. For the construction of the polar code, we use the recursive relation of the reliability sequence of a binary erasure channel (BEC), defined in \cite{ref11}. 
\subsubsection{Performance metrics} Hypothesis testing is used to differentiate Alice from Mallory. Two hypothesis $H_0$ and $H_1$ corresponding respectively to the normal case (Alice) and the spoofing one (Mallory), are defined as follows:
\begin{equation}
    \centering
    \left\{ \begin{aligned}
H_0: \hspace{0.2cm}&\eta=\mathcal{H}_d\left(\mathbf{r}_a(t), \mathbf{r}_a(t+1)\right) \leq \eta_{t h} \\
H_1: \hspace{0.2cm} &\eta=\mathcal{H}_d\left(\mathbf{r}_a(t), \mathbf{r}_m(t+1)\right)>\eta_{t h}
    \end{aligned}
\right.
\label{eq}
\end{equation}
As in \cite{ref_vtc2024}, a bitwise comparison between $\mathbf{r}_a(t)$ and $\mathbf{r}_u(t+1)$, $u \in \{a,m\}$, is a suitable choice for $\eta$. $\mathcal{H}_d(\cdot)$ is then the Hamming distance between $\mathbf{r}_a(t)$ and $\mathbf{r}_u(t+1)$ as (\ref{eta}).
\begin{equation}
\eta = \mathcal{H}_d\left(\mathbf{r}_a(t), \mathbf{r}_u(t+1)\right) = \sum_{j=1}^K\left|r_{a,j}(t)-r_{u,j}(t+1)\right|
\label{eta}
\end{equation}
The main steps of the proposed PLA scheme are summarized in Algorithm \ref{algo_auth}.

\textcolor{black}{
\begin{algorithm}[t]
\caption{Multi-User Physical Layer Authentication}\label{algo_auth}
\begin{algorithmic}[1]
\State \textbf{Enrollment Phase:}
\State Initialization: $N_b$, $\beta$, $U$, $\alpha_i^a$, $\alpha_i^m$, $\forall$ $i = 1, \ldots U$, $M$, $\sigma_z$
\State Measure $\hat{\mathbf{h}}_{a}(t)$ and calculate $\mathbf{x}_a(t)$
\State Quantize $\mathbf{x}_a(t)$ to get $\mathbf{q}_a(t)$
\State Get $\mathbf{r}_a(t)$ (information bits) and frozen bits from $\mathbf{q}_a(t)$  
\State \textbf{Authentication Phase:}
\State Measure $\hat{\mathbf{h}}_{u}(t+1)$ and calculate $\mathbf{x}_u(t+1)$, $u \in \{a,m\}$
\State Quantize $\mathbf{x}_u(t+1)$ to get $\mathbf{q}_u(t+1)$
\State Decode $\mathbf{q}_u(t+1)$ to get $\mathbf{r}_u(t+1)$
\State \textbf{Authentication Decision:}
\State Calculate $\eta = \mathcal{H}_d\left(\mathbf{r}_a(t), \mathbf{r}_u(t+1)\right)$.
\If {$\eta \leq \eta_{th}$}
    \State Authenticate as Alice (Hypothesis $H_0$ is true)
\Else
    \State Identify as Mallory (Hypothesis $H_1$ is true)
\EndIf
\end{algorithmic} 
\end{algorithm}}  
\subsubsection{Algorithm complexity} The time complexity of the enrollment phase is given by $O(N) + O(L) + O(L_snN\log(nN))$. The time complexity of the authentication phase and the decision using hypothesis testing are respectively given by $O(UN) + O(L) + O(L_snN\log(nN))$ and $O(K)$. $L_s$ is the list size and $K$ is the length of the reconciled vectors. The space complexity is the amount of memory required to perform the authentication process taking into account all steps of the scheme is given by $O(U) + O(L) + O(L_snN)$.

\section{Analysis}
In this work, the performance metrics are the probabilities of false alarm and of detection given respectively by $\mathrm{P_{FA}}=\operatorname{Pr}\left(\eta>\eta_{t h} \mid H_0\right)$ and $\mathrm{P_{D}}=\operatorname{Pr}\left(\eta > \eta_{t h} \mid H_1\right)$. Given the probability distribution of $\eta$, we can represent the receiver operating characteristic (ROC) curve. Based on the definition of $\eta$, the probability distributions of $\eta$ are given in Propositions 1 and 2 under $H_0$ and $H_1$, respectively.

\subsubsection*{Proposition 1}
Under $H_0$, $\eta$ follows a binomial distribution of parameters $K$ and $p_0$, i.e. $\eta \sim  \mathbb{B}(K,p_0)$.
\begin{equation}
    \centering
    P(\eta=k | H_0) = \binom{K}{k} p_0^k (1-p_0)^{K-k},
\end{equation}
where $p_0$ is the bit error probability during the decoding.
\subsubsection*{Proposition 2} Under $H_1$, $\eta$ follows a binomial distribution of parameters $K$ and $p_1$, i.e. $\eta \sim  \mathbb{B}(K,p_1)$. 
\begin{equation}
    \centering
P(\eta=k | H_1) = \binom{K}{k} p_1^k (1-p_1)^{K-k},
\end{equation}
where $p_1$ is the bit error probability during the decoding. 

See proofs of Proposition 1 and 2 in \cite{ref_vtc2024}. 

\begin{figure}[t]
     \centering
    \includegraphics[width=0.46\textwidth, height=0.32\textwidth]{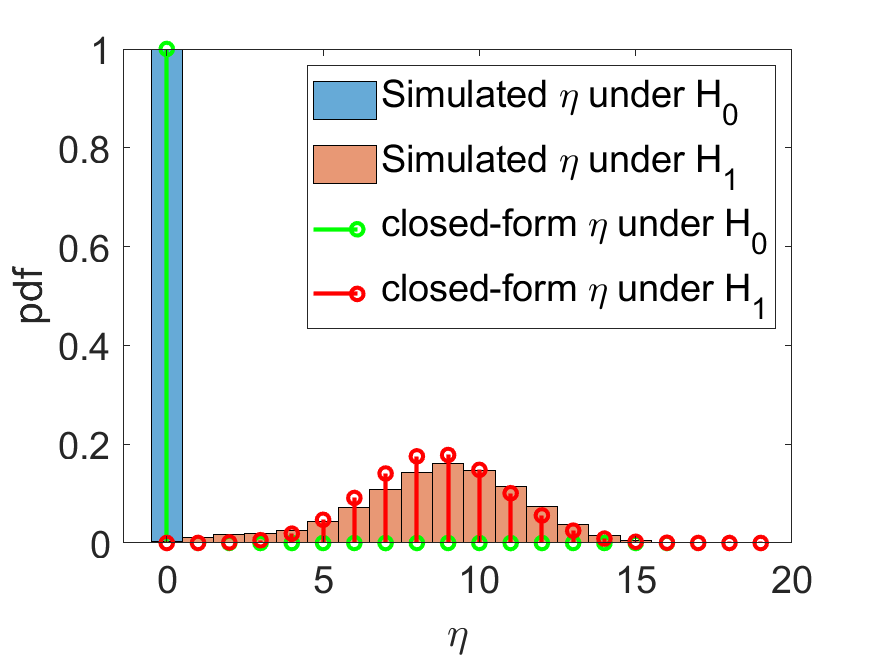}
     \caption{Simulated vs closed-from expression: code rate $ = 0.01$, $SNR = 10~dB$, $p_0 \approx 0$ and $p_1 \approx 0.4539$}
     \label{fig3}
\end{figure}

We compare the simulated probability distributions of $\eta$ with the closed-form ones in the case of a 2-bit quantizer in Fig. \ref{fig3}. The closed-form expression almost matches the simulated result under both hypotheses. As can be seen, the  reconciliation collapses the CSI observations of Alice to a single point probability mass function. As a result, the distance between the probability density function (PDF) of Alice's and Mallory's observations increases. Thus, despite the loss of precision due to the use of quantization, an enhancement of the PLA accuracy is achieved, as confirmed in the numerical results section. 

Given the PDFs of $\eta$, the closed-form expressions of $P_{FA}$ and $P_D$ are respectively given by 
\begin{equation}
    \centering
 P_{FA} = \sum_{k=\eta_{th}+1}^{K} \binom{K}{k} p_0^k (1-p_0)^{K-k}
 \label{equ_pfa}
\end{equation}
and 
\begin{equation}
    \centering
 P_D = \sum_{k=\eta_{th}+1}^{K} \binom{K}{k} p_1^k (1-p_1)^{K-k}.
 \label{equ_pd}
\end{equation}

\section{Numerical Results}

We compare here the proposed method with the prior ones proposed in \cite{ref6}, \cite{ref7} and \cite{phase}. We first investigate the ROC curve which is $P_D$ vs $P_{FA}$ and the impact of the $\mathrm{SNR}$ on the probability of detection. Then, we provide the analysis of the other simulation setup parameters such as the correlation coefficient $\beta$, the interference weights $\alpha_i^a$ and $\alpha_i^m$ and the code rate. We also present the results of the proposed method for different codelengths, $nN = 1024$ and $nN = 2048$, which correspond to 1-bit ($n = 1$) and 2-bit ($n = 2$) quantizers, respectively. 

Unless otherwise specified, the simulation parameters in this work are considered as follow. The number of antennas at Bob $N_b = 32$, $\beta = 0.9$, $\sigma_h^2 = 1$, $M = 16$, $N = 1024$, the code rate is $0.01$, $\alpha_i^a = \alpha_i^m = 0.01$, $\forall$ $i$ and $P_{FA} = 10^{-3}$. This very small value of $P_{FA}$ is chosen in order to match real-world communication systems in which low or very low false alarm probabilities are needed. The signal-to-noise ratio is defined as $SNR = \frac{\mathbb{E}[\mathbf{h}_a\mathbf{h}_a^H]}{N_b\sigma_z^2}$.

\begin{figure}[t]
\centering
\includegraphics[width=0.4\textwidth, height=0.25\textwidth]{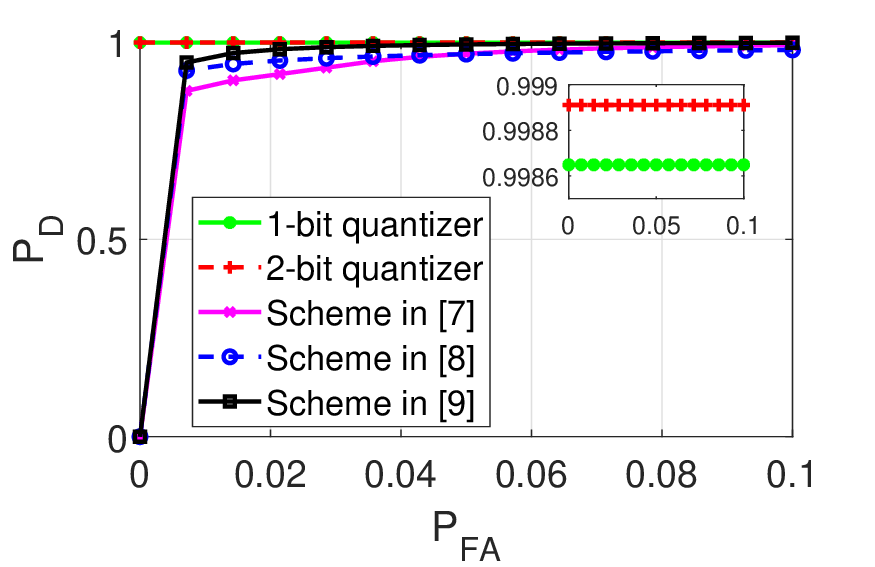}
\vspace{-0.5em}
\caption{ROC curve: $SNR = 5~dB$}
\label{fig_roc}    
\end{figure}

Fig. \ref{fig_roc} presents the $\mathrm{ROC}$ curve for a SNR of $5~dB$. Our proposed method performs very well with a probability of detection very close to $1$ for very small probabilities of false alarm less that $0.1$. It also performs better than the prior schemes. As shown in this figure, we get more than $99.80\%$ increase in the detection probability even for false alarm probabilities very close to $0$.

\begin{figure}[t]
\centering
\includegraphics[width=0.4\textwidth, height=0.25\textwidth]{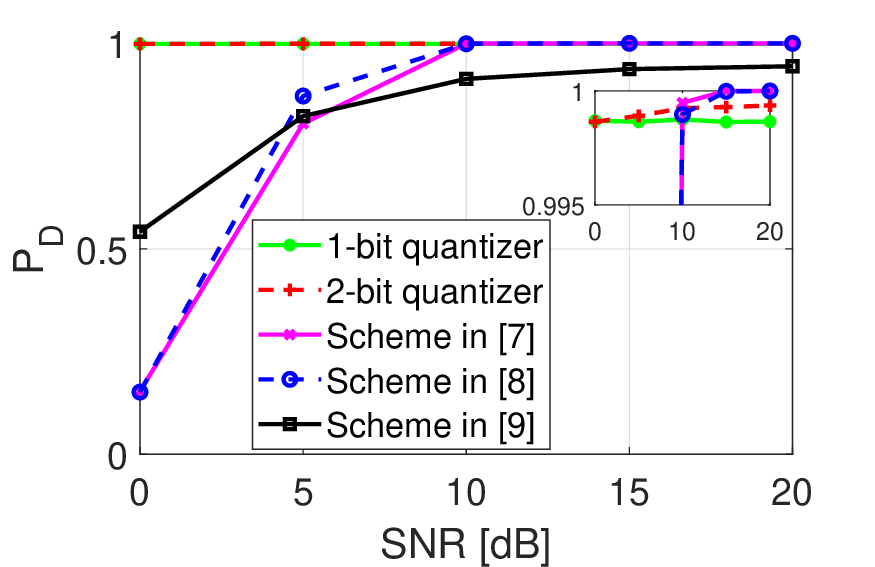}  
\vspace{-0.5em}
\caption{$P_D$ vs $SNR$: $P_{FA} = 10^{-3}$ }
\label{fig_snr}    
\end{figure}

In Fig. \ref{fig_snr}, the impact of the $\mathrm{SNR}$ on the detection probability is studied. As the SNR increases, $P_D$ increases as subsequent CSI measurements are more correlated due to decreased noise. The proposed scheme performs very well with a $P_D$ greater than $99.86\%$. However the performance of the prior methods is poor for low SNRs less than $10~dB$.

\begin{figure}[t]
\centering
\includegraphics[width=0.4\textwidth, height=0.25\textwidth]{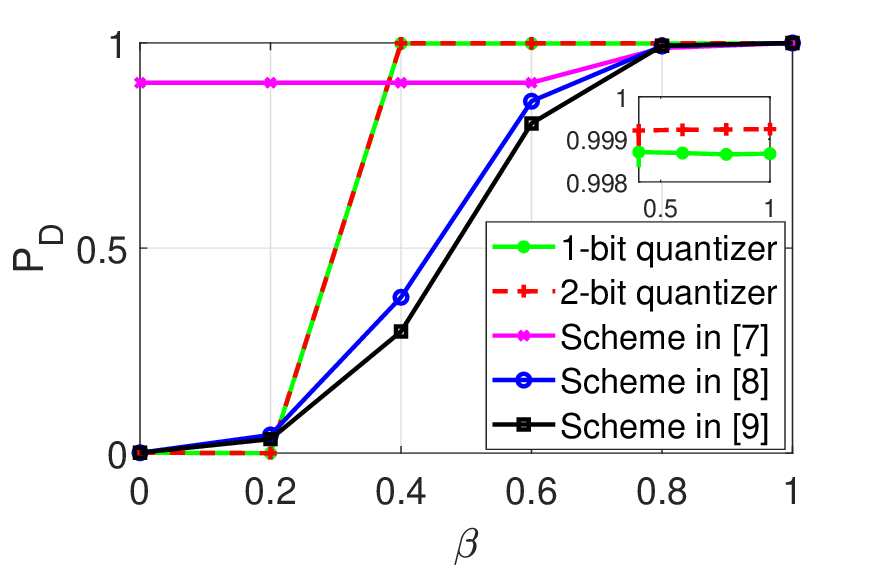}
\vspace{-0.5em}
\caption{$P_D$ vs $\beta$: $SNR = 10~dB$, $P_{FA} = 10^{-3}$}
\label{fig_beta}    
\end{figure}

Fig. \ref{fig_beta} shows the detection probability as a function of $\beta$. We have a $P_D$ very close to $1$ for $0.4 \leq \beta \leq 1$. We therefore have a very well performance even for poor scenarios of medium correlation coefficients. Thus, our proposed scheme performs better than the previous ones. We observe that the scheme in \cite{ref6} is performing always good even for $\beta$ less than $0.4$. Actually, \cite{ref6} proposed a two dimensional authentication scheme where the channel impulse response and the multipath delay are considered. This performance is due to the high correlations in the multipath delay dimension.

\begin{figure}[t]
\centering
\includegraphics[width=0.4\textwidth, height=0.266\textwidth]{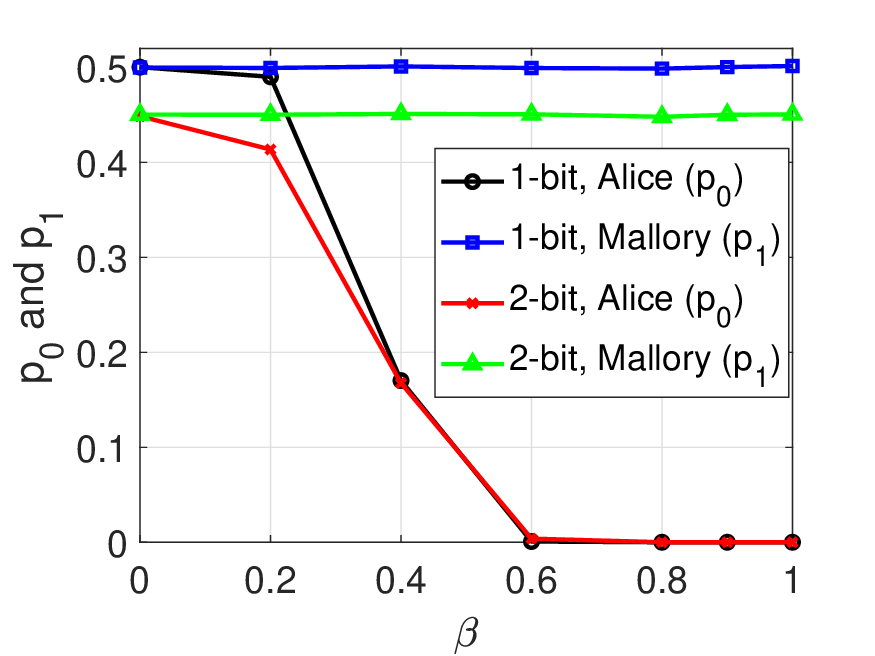}
\vspace{-0.5em}
\caption{Error probabilities vs $\beta$: $SNR = 5~dB$}
\label{fig_proba}    
\end{figure}
Table \ref{table_alpha} presents the behaviour of $P_D$ for different values of $\alpha^a = \alpha^m$ under a $SNR = 10~dB$. We found that, for the given system parameters, the maximum interference weight that achieves almost perfect reconciliation is equal to $0.8$ for the proposed scheme. The techniques reported in \cite{ref7} and \cite{phase} perform poorly from $\alpha^a = \alpha^m = 0.8$, that is, the detection probabilities are respectively equal to $0.22$ and $0.43$ for $\alpha^a = \alpha^m = 0.8$. The work in \cite{ref6} has a good performance for weights above $1.6$ because of the high correlation in the delays as mentioned before.
\begin{table}[h!]
\centering
\caption{Detection probabilities for different $\alpha^a = \alpha^m$} 
\begin{tabular}{|c|c|c|c|c|c|}
\hline
$\alpha^a = \alpha^m$ & 1-bit & 2-bit & in \cite{ref6} & in \cite{ref7} & in \cite{phase} \\ \hline
0                     & 0.99     & 0.99     & 0.99          & 0.99           & 0.99           \\ \hline
0.01                  & 0.99     & 0.99     & 0.99          & 0.99           & 0.91           \\ \hline
0.8                   & 0.99     & 0.99     & 0.90       & 0.22        & 0.43        \\ \hline
1.6                   & 0     & 0     & 0.90       & 0.01        & 0.07        \\ \hline
2                     & 0     & 0     & 0.90       & 0.005       & 0.04        \\ \hline
\end{tabular}
\label{table_alpha}
\end{table}

Error probabilities $p_0$ and $p_1$ mentioned in Section IV are evaluated in Fig. \ref{fig_proba} for a false alarm of $10^{-3}$. For both 1-bit and 2-bit quantizers, the normal case ($H_0$) is performing better the spoofing case ($H_1$). This can be explained by the correlation between the channel measurements in the normal case.

An analysis of the code rate of the proposed scheme is shown in Table \ref{table_rate} for a SNR of $10~dB$. We can see that the performance is almost perfect for a code rate of $0.01$ and poor for values greater than $0.01$. It can be improved for code rates greater than $0.01$ by increasing the code length or designing an improved coding scheme.
\begin{table}[h!]
\centering
\caption{Detection probabilities under different code rates}  
\begin{tabular}{|c|c|c|c|c|c|}
\hline
Code  rate & 0.01 & 0.1 & 0.2 & 0.3 & 0.4 \\ \hline
1-bit      & 0.998    & 0   & 0   & 0   & 0   \\ \hline
2-bit      & 0.999    & 0   & 0   & 0   & 0   \\ \hline
\end{tabular}
\label{table_rate}
\end{table}

All the previous presented results are shown for 1-bit and 2-bit quantizers. The 2-bit quantizer performs better than the 1-bit one as there is more bits that increase the code length. This also shows the impact of the code length on the performance.
\section{Conclusion}
In this work, the problem of CSI-based PLA using information reconciliation has been investigated. An information reconciliation scheme using Polar codes along with a quantization strategy that employs an arbitrary number of bits has been developed for a multiuser communication scenario. CSI measurements from a legitimate user spread in time are reconciled using reconciliation decoders implemented using Polar codes. Simulation results have shown that our approach performs excellently well in the presence of multiple users and outperforms state-of-the art schemes. 

\section*{Acknowledgment}
A. K. A. Passah has been supported by the project SRV ENSEA. A. Chorti has been partially supported by the EC through the Horizon Europe/JU SNS project ROBUST-6G (Grant Agreement no. 101139068), the IPAL project CONNECTING, the ANR-PEPR 5G Future Networks project and the CYU INEX-PHEBE project. R. C. de Lamare has been supported by PUC-Rio, CNPq, FAPERJ and FAPESP.


\begin{thebibliography}{00}
\bibitem{ref_IoT}Z. Lin, M. Lin, T. de Cola, J. -B. Wang, W. -P. Zhu and J. Cheng, "Supporting IoT With Rate-Splitting Multiple Access in Satellite and Aerial-Integrated Networks," in \textit{IEEE Internet of Things Journal,} vol. 8, no. 14, pp. 11123-11134, July, 2021.
\bibitem{ref3}Shakiba-Herfeh, M., Chorti, A., Vincent Poor, H. (2021). \textit{Physical Layer Security: Authentication, Integrity, and Confidentiality.} In: Le, K.N. (eds) Physical Layer Security. Springer, Cham. 
\bibitem{ref4}M. Mitev, A. Chorti, H. V. Poor and G. P. Fettweis, "What Physical Layer Security Can Do for 6G Security," in \textit{IEEE Open Journal of Vehicular Technology,} vol. 4, pp. 375-388, 2023. 
\bibitem{ref1}Bloch, M., and Barros, J. (2011). \textit{Physical-Layer Security: From Information Theory to Security Engineering.} Cambridge: Cambridge University Press. 
\bibitem{srmax}X. Lu and R. C. de Lamare, "Opportunistic Relaying and Jamming Based on Secrecy-Rate Maximization for Multiuser Buffer-Aided Relay Systems," in \textit{IEEE Transactions on Vehicular Technology,} vol. 69, no. 12, pp. 15269-15283, Dec. 2020.
\bibitem{ref__}Y. Shi, X. Lu, K. An, Y. Li and G. Zheng, "Efficient Index-Modulation-Based FHSS: A Unified Anti-Jamming Perspective," in \textit{IEEE Internet of Things Journal,} vol. 11, no. 2, pp. 3458-3472, July 2023.
\bibitem{ref6}J. Liu and X. Wang, "Physical Layer Authentication Enhancement Using Two-Dimensional Channel Quantization," in \textit{IEEE Transactions on Wireless Communications,} vol. 15, no. 6, pp. 4171-4182, June 2016. 
\bibitem{ref7}N. Xie, J. Chen and L. Huang, "Physical-Layer Authentication Using Multiple Channel-Based Features," in \textit{IEEE Transactions on Information Forensics and Security,} vol. 16, pp. 2356-2366, 2021. 
\bibitem{phase}X. Lu, J. Lei, Y. Shi and W. Li, "Physical-Layer Authentication Based on Channel Phase Responses for Multi-Carriers Transmission," in \textit{IEEE Transactions on Information Forensics and Security,} vol. 18, pp. 1734-1748, 2023.
\bibitem{ref_vtc2024}A. K. A. Passah, R. C. de Lamare and A. Chorti, "Physical Layer Authentication Using Information Reconciliation", \textit{2024 IEEE 99th Vehicular Technology Conference (VTC2024-Spring)}, Singapore, 2024.
\bibitem{ref_prec}T. M. Pham, L. Senigagliesi, M. Baldi, G. P. Fettweis and A. Chorti, "Machine Learning-based Robust Physical Layer Authentication Using Angle of Arrival Estimation," in \textit{IEEE Global Communications Conference (GLOBECOM),} Dec 2023, Kuala Lampur, Malaysia.
\bibitem{jakes}Jakes William C., "Microwave Mobile Communications," \textit{IEEE,} 1994.
\bibitem{goldsmith}A. Goldsmith, "Wireless communications," \textit{Cambridge university press,} 2005.
\bibitem{ref8}M. Shakiba-Herfeh and A. Chorti, "Comparison of Short Blocklength Slepian-Wolf Coding for Key Reconciliation," \textit{2021 IEEE Statistical Signal Processing Workshop (SSP),} Rio de Janeiro, 2021, pp. 111-115. 
\bibitem{pilot}H. N. Qureshi, I. H. Naqvi and M. Uppal, "Massive MIMO with Quasi Orthogonal Pilots: A Flexible Solution for TDD Systems," \textit{2017 IEEE 86th Vehicular Technology Conference (VTC-Fall),} Toronto, ON, Canada, 2017, pp. 1-6.
\bibitem{ref9}E. Arikan, "Source polarization," \textit{2010 IEEE International Symposium on Information Theory,} Austin, TX, USA, 2010, pp. 899-903. 
\bibitem{ref10}I. Tal and A. Vardy, "List decoding of polar codes," \textit{2011 IEEE International Symposium on Information Theory Proceedings,} St. Petersburg, Russia, 2011, pp. 1-5. 
\bibitem{ref11}E. Arikan, "Channel Polarization: A Method for Constructing Capacity-Achieving Codes for Symmetric Binary-Input Memoryless Channels," in \textit{IEEE Transactions on Information Theory,} vol. 55, no. 7, pp. 3051-3073, July 2009.  
\end{thebibliography}
\end{document}